\documentclass[useAMS,usenatbib]{mn2e}

\usepackage{epsfig}
\title[Evolution of blue supergiants] %and $\alpha$ Cygni variables]
{Evolution of blue supergiants and $\alpha$ Cygni variables;
Puzzling CNO surface abundances}
\author[H. Saio, C. Georgy, and G. Meynet]{Hideyuki Saio$^{1}$\thanks{E-mail:
saio@astr.tohoku.ac.jp},  Cyril Georgy$^{2,3}$, and Georges Meynet$^{4}$\\
$^{1}$Astronomical Institute, Graduate School of Science, Tohoku University, Sendai, Japan\\
$^{2}$Centre de Recherche Astrophysique de Lyon, \'Ecole Normale Sup\'erieure de Lyon, 46, all\'ee d$^,$Italie, 69384 Lyon Cedex 07, France\\
$^{3}$Astrophysics, Lennard-Jones Laboratories, EPSAM, Keele University, ST5 5BG, Staffordshire, UK \\
$^{4}$Geneva Observatory, University of Geneva, Maillettes 51, 1290 Sauverny, Switzerland
}

\def\bm#1{\mbox{\boldmath$#1$}}

\begin{document}

\date{}

\pagerange{\pageref{firstpage}--\pageref{lastpage}} \pubyear{2013}

\maketitle

\label{firstpage}

\begin{abstract}
A massive star can enter the blue supergiant region either evolving 
directly from the main-sequence,  or evolving from a previous 
red supergiant stage. 
The fractions of the blue supergiants having different histories depend on the
internal mixing and mass-loss during the red supergiant stage.
We study the possibility to use diagnostics based on stellar pulsation to
discriminate blue supergiants having different evolution histories.
For this purpose we have studied the pulsation property of massive star models
calculated with the Geneva stellar evolution code for 
initial masses ranging from 8 to 50\,M$_\odot$ with a 
solar metallicity of $Z=0.014$.
We have found that radial pulsations are excited in the blue-supergiant
region only in the models that had been red-supergiants before. 
This would provide us with a useful mean to diagnose the history 
of evolution of each blue-supergiant. 
At a given effective temperature, much more nonradial pulsations are 
excited in the model after the red-supergiant stage than 
in the model evolving towards the red-supergiant.  
The properties of radial and nonradial pulsations in 
blue supergiants are discussed. 
Predicted periods are compared with period ranges observed in 
some $\alpha$-Cygni variables in the Galaxy and NGC\,300.
We have found that blue supergiant models 
after the red-supergiant stage 
roughly agree with observed period ranges in most cases.
However, we are left with the puzzle that the 
predicted surface N/C and N/O ratios seem 
to be too high compared with those of Deneb and Rigel.
\end{abstract}


\section{Introduction}
The post-main-sequence evolution of massive stars depends sensitively on 
the helium core mass and its ratio to the envelope mass, which in turn 
depends on still poorly understood phenomena such as mixings
in the radiative layers (core overshooting and rotational mixing) 
and wind mass loss.
Recent evolution models with a solar metallicity of $Z=0.014$ by \citet{eks12}
indicate that a star with a sufficiently large initial mass undergoes
a blue-red-blue (or blue-loop) evolution before central helium exhaustion;
i.e., the star ignites He in the center in the blue supergiant (BSG) stage,  
evolves to the red-supergiant (RSG) region,  and returns to
the blue supergiant (BSG) region during core He-burning. 
The lowest initial-mass for the blue-red-blue evolution depends on
the degree of mixing in radiative layers and the strength of wind mass loss.
\citet{eks12}'s results indicate the lower bound to be about  20\,M$_\odot$. 
The mass limit is lowered  
if higher mass-loss rates in the RSG phase is assumed 
\citep{geo12,sal99,vanb98}.

Thus, luminous BSGs consist of two groups having different evolution histories: 
one group are evolving red-wards just after the termination of main-sequence, 
while another group have  evolved back from the RSG stage.
The BSGs belonging to the latter group have significantly reduced envelope 
mass and the surface is contaminated by the CNO-processed matter 
due to a dredge-up in the RSG stage and a significant mass loss.
The fraction of each group depends on
the internal mixing in the radiative layers and 
the strength of stellar wind and metallicity.
In other words, if we can distinguish the two kinds of BSGs, it would be
very useful for constraining the mixing in radiative layers and wind parameters.
Furthermore, the fraction relates to the relative frequencies of different
types of core-collapse supernovae such as IIP, IIL, IIb, Ib and Ic
\citep[e.g.,][]{geo12b,yoo12,eld13,vanb12}
and the ratio of blue to red supergiants \citep[e.g.,][]{mey83,lan95,egg02}.

One way to distinguish the two groups
is to obtain their surface abundances of the CNO elements. 
This has been pursued intensively by many authors; e.g., the VLT-FLAME survey
\citep{hun09}, \citet{prz06} and \citet{tak00}.
Although the majority of BSGs show enhanced N/C ratios, 
theoretical interpretations 
were somewhat hampered by the variety of rotation velocities which
yield various degree of internal mixings in the main-sequence stage, 
and possible effect of close binaries and magnetic fields.

We propose, in this paper, another way to distinguish the two groups of 
BSGs by using stellar pulsation; i.e., we will argue that if they show 
(radial) pulsations, they must have been red supergiants before. 
It is known that many luminous ($\log L/{\rm L}_\odot \ga 4.6$) BA-supergiants 
in our Galaxy and Magellanic Clouds show micro variations in luminosity 
and in radial velocities; they are called $\alpha$-Cygni variables 
\citep[e.g., ][]{van98}.
In addition, \citet{bre04} found that a fraction of blue supergiants 
in the galaxy NGC\,300 are such
variables and at least two of those show clear radial pulsation properties.
The NGC\,300 BSGs would be particularly useful
for constraining evolutionary models, because of the homogeneity of the
data and less ambiguities in luminosity.

The pulsation not only provides us with diagnostic means, it might also have
effects on stellar winds from massive stars, 
as \citet{aer10b} found a relation between episodic changes in mass loss
and the 37\,day pulsation of the luminous blue supergiant HD 50064. 
They suggested that the pulsation is a radial strange-mode pulsation,
which we confirm in this paper.

The paper is organized as follows: evolution models of massive stars and the 
excitation of radial pulsations in these models are discussed in \S2.
The properties of radial and nonradial pulsations and their excitation mechanisms
are discussed in \S3.  In \S4 we compare observed semi-periods of $\alpha$-Cygni
variables with theoretical ones and discuss  surface compositions.
Our conclusion is given in \S5.

\section{Massive star evolution and the stability to radial pulsations}
Evolutionary models have been calculated by the Geneva evolution code with 
the same input physics as those described in \citet{eks12}.
The initial abundances adopted are $(X,Z)=(0.720, 0.014)$ with a solar mixture
for the heavy elements \citep[][for the Ne abundance]{asp05,cun06}.
A core overshooting of 0.1 pressure scale height is included.
Stellar mass loss rate for a given position on the HR diagram and current
mass is obtained from the prescriptions described  in \citet{eks12}
(Except for $14\,M_\odot$ models, see below).

\begin{figure} %fig1
\epsfig{file=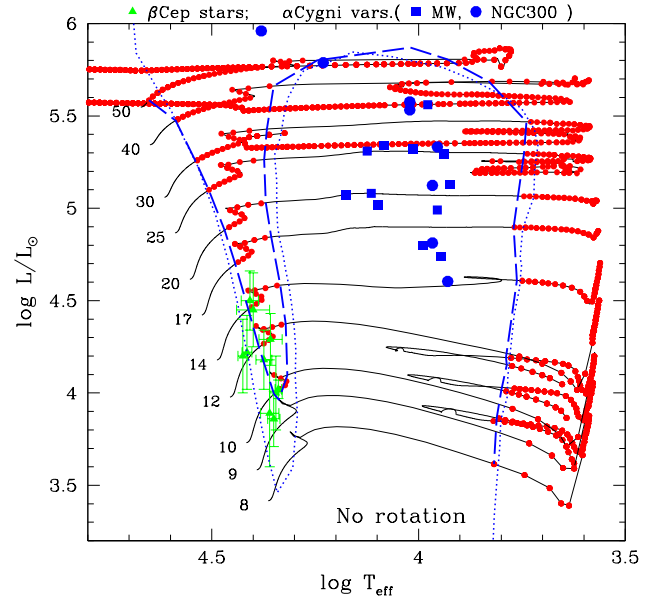,width=0.49\textwidth}
\caption{Evolutionary tracks for massive models without rotation 
for a chemical composition of $(X,Z)=(0.72,0.014)$ and
the stability boundary (dashed line) for low-order radial pulsations.
For models with $M_{\rm i} \ge 25$\,M$_\odot$ the stability boundary 
is determined for models evolving red-ward for the first time.
Dotted line indicates the stability boundary for $(X,Z)=(0.70,0.02)$
\citep[GN93 abundance][]{gre93}.
Red dots along evolutionary tracks indicate the positions of the models in which
at least one radial modes is excited.
Also plotted are some of the observed $\alpha$ Cygni variables 
(filled circles and squares) and
$\beta$ Cephei variables (triangles with error bars).
For the parameters of the Galactic $\alpha$ Cygni variables, we referred to
various literature including that listed in Table\,\ref{tab:param}, and
\citet{kud99}. 
For $\alpha$ Cygni variables in NGC\,300 we referred to
\citet{bre04} and \citet{kud08}.  
Parameters adopted for $\beta$ Cephei variables are listed in Table\,\ref{tab:betcep}.
Small jumps during the Cepehid loops for  $M_{\rm i}\le 10M_\odot$
are produced by quick extensions of the convective core.
}
\label{fig:stb}
\end{figure}

\begin{table}
\caption{Parameters of $\beta$ Cephei variables$^*$ adopted in Fig.\,\ref{fig:stb}}
\begin{tabular}{@{}lccccc@{}}
%# parameters of beta Cephei variables
\hline
Name  &   $\log T_{\rm eff}$ &  $\pm$  &   $\log L/L_\odot$ &  $\pm$ & ref  \cr
\hline
15 CMa    &   4.408  & 0.021  & 4.50   & 0.16 & a \cr %  Shobbrook et al (2006)
$\beta$ CMa &  4.40 &  0.04  &  4.45  & 0.20  & b \cr % Mazumdar et al (2006)
BW Vul   &  4.358  & 0.028  &  4.29  & 0.14  & c \cr % Fokin et al (2004)   
KZ Mus  &   4.415  & 0.012  &  4.22  & 0.20  & d \cr % Handler et al (2003)
V433 Car  &  4.425 & 0.012  & 4.20  & 0.2  & d \cr %  Hander et al (2003)
%IL Vel   &  4.373  & 0.011  &  4.19  & 0.22  & d  \cr  % Hander et al (2003)
12 Lac  &   4.374  & 0.019  & 4.18 & 0.16  & e  \cr  % Handler et al (2006)
$\delta$ Cet & 4.339 & 0.008 &  4.02 & 0.05 & f \cr % Aerts et al (2006)
$\nu$ Eri   &  4.360  & 0.022  &  3.89  & 0.29 &  g \cr % De Ridder et al (2004)
%$\nu$ Eri   &  4.346  & 0.011  &  3.94  & 0.15 &  g \cr % pamyatnykh et al (2004)
16 Lac & 4.345 & 0.015 &  4.0 &  0.2 & h \cr %   Thoul et al (2003)
HD129929 &  4.350 &  0.015  &  3.86 & 0.15 & i \cr %  Aerts et al(2004) Dupre et al (2004)
\hline
\end{tabular}
\label{tab:betcep}
a=\citet{sho06}, b=\citet{maz06}, c=\citet{fok04}, d=\citet{han03},
e=\citet{han06}, f=\citet{aer06}, g=\citet{der04} %g=\citet{pam04}
, h=\citet{tho03},
i=\citet{dup04}\\
$^*$This is a very incomplete sample of Galactic $\beta$ Cep variables 
collected only for illustrative purpose in Fig.\,\ref{fig:stb}.
\end{table}

\begin{figure} %fig2
\epsfig{file=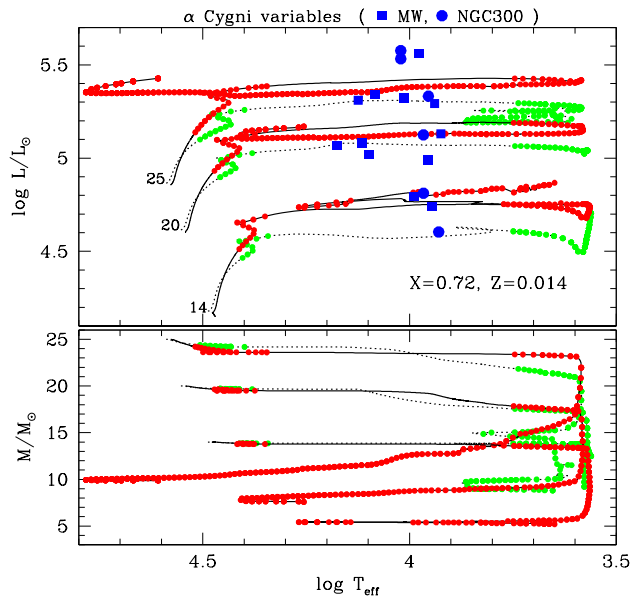,width=0.49\textwidth}
\caption{Top panel shows evolutionary tracks for 
$M_{\rm i}= 25, 20$, and 14\,M$_\odot$, and the bottom panel shows 
the decrease of mass during the evolution for each initial mass.
Solid and dotted lines are for models with  and without rotation, respectively.
Rotation rates are assumed to be 40\% of the critical rate at the zero-age 
main-sequence stage. 
Red and green dots along each line indicate models 
in which at least one radial pulsation mode is excited.
Some of the $\alpha$ Cygni variables in our Galaxy 
(filled squares) and in the galaxy NGC\,300 (filled circles) are also plotted.
}
\label{fig:mass}
\end{figure}

Fig.\,\ref{fig:stb} shows evolutionary tracks up to the central helium 
exhaustion calculated without including rotational mixing for initial masses 
of 8, 9, 10, 12, 14, 17, 20, 25, 30, 40, and 50\,M$_\odot$.
For $M_{\rm i}\ge 12$\,M$_\odot$, the helium burning starts when stars are
evolving in the blue supergiant (BSG) region after the termination of 
main-sequence stage. 
As He burns in the center, they evolve into the red supergiant (RSG) stage. 
Stars with $M_{\rm i} \ge 30$\,M$_\odot$ evolve back to the BSG region
(blue-loop) before the helium is exhausted in the center.
A star starts a blue-loop when it loses enough mass in the RSG stage
(Fig.\,\ref{fig:mass}). 
This has an important consequence for the stability of radial pulsations.

We have performed a stability analysis of radial pulsations 
for selected evolutionary models. The method is
described in \citet{sai83}, where the perturbation of the divergence of 
convective flux and the effect of rotation is neglected.
The latter is justified because
rotation is always slow in the envelope of supergiants
(this is also true in our models with rotation; see Appendix),
where pulsations have appreciable amplitudes.
The effect of convection is neglected because the theory for 
the convection-pulsation coupling is still infant.
Since the convective flux is less than 50\% of the total flux 
in the convection zones in the envelope
of blue supergiants, we do not think that neglecting the convection-pulsation
coupling affects significantly our results. 

The outer boundary is set at an optical depth of $\sim10^{-4}$.
We have adopted the outer mechanical condition that
the Lagrangian perturbation of the gas pressure goes to zero.

The red dots along evolutionary tracks in Fig.\,\ref{fig:stb} indicate the positions 
of the models that are found to have at least one excited radial mode.
The dashed line in Fig.\,\ref{fig:stb} indicates the stability 
boundary of radial low-order pulsations, which are appropriate 
for the Cepheids and $\beta$ Cephei variables.  
For models with $M_{\rm i}\ge 30$\,M$_\odot$, which make blue-red-blue evolution,
the part evolving toward the RSG region (first crossing) was used to obtain
the stability boundary.
For comparison, the stability boundary for models with the abundance
$Z=0.02$ \citep{sai11} with GN93 mixture \citep[][]{gre93} is also shown by 
a dotted line.  

The nearly vertical `finger' of the instability boundary 
around $\log T_{\rm eff}\sim 4.3 - 4.4$ corresponds 
to the $\beta$ Cephei instability region
(excited by the $\kappa$-mechanism at the 
Fe-opacity bump around $T\sim2\times10^5$K), 
while the vertical boundary at $\log T_{\rm eff}\sim3.8$
is the blue edge of the Cepheid instability strip, in which pulsations are 
excited at the second helium ionization zone. 
(No red-edge is obtained because our pulsation analysis does not include 
the coupling between pulsation and convection.)

The boundary for the $\beta$ Cephei  instability 
region depends on the metal abundance.
The positions of some Galactic
$\beta$ Cephei stars are shown in Fig.\,\ref{fig:stb} 
by filled triangles with error bars.
Comparing the distribution of the $\beta$ Cephei variables with
the stability boundaries for $Z=0.014$ (dashed line) and 
$Z=0.02$ (dotted line), we see that  
the most appropriate heavy-element abundance for the Galactic
$\beta$ Cephei variables seems to be slightly larger than $Z=0.014$.
The other part of the instability boundary hardly depends on the metallicity.

The instability boundary for radial pulsations 
by (weakly non-adiabatic) $\kappa$-mechanisms have steep gradients
in the HR diagram as seen in the less luminous part ($\log L/{\rm L}_\odot \la 5$)
of Fig.\,\ref{fig:stb},
where the instability boundaries for $Z=0.014$ and 0.02 are shown by 
broken lines.
This comes from the requirement that
the pulsation period should be comparable to the thermal timescale 
at the zone where the $\kappa$-mechanism works \citep[e.g.,][]{cox74}.
At high luminosity, the instability boundary is nearly horizontal.
This is associated with  the strange modes that occur 
if $L/M \ga 10^4\,{\rm L}_\odot/{\rm M}_\odot$ 
\citep[e.g.,][and discussion below]{gau90, gla94, sai98}.

In the BSG models evolving toward the RSG stage 
no radial modes are excited between the 
red-boundary of the $\beta$ Cephei instability region and the Cepheid 
blue-edge, because $L/M$ is not sufficiently large for the strange mode
mechanism to work.
Models  with $M_{\rm i} \ge 30$\,M$_\odot$ return to the BSG region 
(blue-loop) from the RSG region before core-helium exhaustion.
Radial pulsations are excited in the models on the blue-loop; this is
due to the fact that a significant mass is lost in the RSG stage 
and hence the $L/M$ ratio has increased considerably (Fig.\,\ref{fig:mass}). 
We can identify $\alpha$ Cygni variables 
(especially if radial pulsations are involved) 
to be core-helium burning stars
on the blue-loop returned from the RSG stage.
However, the luminosity of the track for $M_{\rm i}=30\,{\rm M}_\odot$ is still too high to
be consistent with the distribution of 
$\alpha$ Cygni variables on the HR diagram.
The discrepancy can be solved by taking into account rotational mixing
\citep{eks12},
or assuming a strong mass loss caused by Roch-Lobe overflow 
in the RSG stage.
We consider the effect of rotational mixing in this paper. 

We have calculated evolution models for $M_{\rm i}=25, 20$, and 14\,M$_\odot$
\footnote{The last model was computed with an increased mass-loss rate
compared to the standard prescription (see below).}
with rotational mixing,
and examined the stability of radial pulsations for those models.
The ways to treat rotation and the mixing are the same as \citet{eks12};
the rotation speed was assumed to be 40\% of the critical one at the zero-age
main-sequence stage.
The results are shown in Fig.\,\ref{fig:mass}.
The rotational mixing makes helium core and hence luminosity larger in the
post main-sequence evolution for a given initial mass.
A smaller ratio of the envelope to core mass makes the red-ward 
evolution faster; i.e., less He is consumed in the first BSG stage. 
(Note that dotted lines %(without rotation) 
in the bottom panel of Fig.\,\ref{fig:mass} tend to
lose more mass as a function of $T_{\rm eff}$ in the first crossing, 
indicating the evolution to be slow there without rotational mixing).
Also, the higher luminosity enhances stellar winds so that the star 
starts blue loop earlier, well before the central He exhaustion.
We note that the important effect of rotation comes from the mixing 
that enlarges He core, but not from the centrifugal force.
Therefore, a similar evolution is possible even without including rotation
if a more extensive core overshooting is assumed.
For $M_{\rm i}= 20\,M_\odot$, for example, an extensive blue-loop
occurs before He exhaustion if a core overshooting larger than 
$\sim 0.4\,H_p$ is included without rotational mixings;
if a mass-loss rate is enhanced by a factor of 5, for example, it 
occurs for a overshooting larger than $\sim 0.35\,H_p$.

The non-rotating evolutionary track of  $M_{\rm i}=14\,$M$_\odot$ passes, 
in the first crossing, 
around the lower bound of the distribution of the $\alpha$ Cygni variables
($\log L/{\rm L}_\odot \ga 4.6$).
However, it does not come back to blue region even if rotation is included
with our standard parameters.
It does make a blue loop as shown in Fig.\,\ref{fig:mass}, 
if the rate of cool winds is increased by a factor of five as in \citet{geo12}. 
Such an increase is reasonable since there are many %arguments both 
theoretical and observational arguments of sustaining higher mass loss rates during the red supergiant phase. 
From a theoretical point of view, \citet{yoo10} %Yoon \& Cantiello (2010) 
have %for instance 
studied the consequence of a pulsation driven mass loss during the red supergiant phase. They showed that using empirical relations between the pulsation period and the mass loss rates, the mass loss rates could be increased by quite large factors largely exceeding a factor 5 at least during short periods. 

From an observational point of view, %the study of 
the circumstellar environment of red supergiants clearly indicates that some stars undergo strong mass loss outbursts. 
For instance, VY CMA (M2.5-5Iae) which has a current mass loss rate of 
2-4 10$^{-4}$M$_\odot$ per year \citep{dan94} %(Danchi et al. 2004) 
is surrounded by a very inhomogeneous nebula likely resulted from a series of episodic mass ejections over the last 1000 yr  \citep{smi09}. %(Smith et al. 2009). 
It is estimated that the mass loss rates between a few hundred and 1000 yr ago was 1-2 10$^{-3}$M$_\odot$ per year, thus between 2.5 and 10 times greater than the present
rate.

We also note that  mass loss rates obtained by \citet{vanL05} for dust enshrouded red 
supergiants are larger by a factor up to 10 compared with the rates estimated 
from the empirical relations given in \citet{dej88} %for typically a 14 M$_\odot$ red supergiant  than the mass loss rates obtained by . %de Jager et al. (1988).

In the present standard computation we used the prescription by \citet{dej88}. %de Jager et al (1988). 
The arguments above indicate that using mass loss rates increased by a factor 5 is not beyond %reasonable 
the uncertainty in our present state of knowledge on the mass loss rates of 
red supergiants.

It is interesting to note that for the case of 14\,M$_\odot$  
not all models on the blue-loop  excite pulsations.
More precisely, no radial modes  are excited in the blueward evolution
at $\log T_{\rm eff}\approx4$ ($\log L/M = 4.03$).
Only in the second
redward evolution (third crossing), a radial mode is excited 
around similar effective temperature; this time, models have 
slightly higher $L/M$ ($\log L/M = 4.08$).

The fact that no $\alpha$ Cygni variables are observed below a 
luminosity limit of about $\log L/{\rm L}_\odot\approx4.6$ does 
not necessarily means that stars 
below that limit have no blue-loop. 
Even if they make a blue-loop evolution, 
their $L/M$ ratio would be too small for pulsations to be excited.

The observed properties of $\alpha$ Cygni variables can be well explained 
if these stars are core He-burning stars on the blue-loop.
This supports the presence of considerable mixing and possibly 
cool winds stronger than adopted in \citet{eks12}. 
Note that an increased mass loss during the RSG phase seems to be also 
required in order to reproduce the observed positions of 
$\alpha$ Cyg variables at high luminosity. 
(Indeed, the models in \citet{eks12} have a mass-loss rate increased 
by a factor of 3 for models more massive than 20\,M$_\odot$ 
during the RSG phase).

From the ratio of the evolution speeds between 
the first (red-ward) and second (blue-ward) crossings we can estimate the
probability for a BSG to be on the first crossing.
At positions of typical BSGs,  Rigel  and Deneb 
(see Table\,\ref{tab:param} below),
for example, the probabilities are  45\% and 98\% , respectively, 
for $M_i=20\,M_\odot$, while they are 15\% and 50\% or $M_i=25\,M_\odot$;
in both cases rotational mixings are included. 
If we use models without rotation, the probabilities are nearly unity for 
both stars and for both initial masses,
because in this case the second crossings occur very swiftly after the core He 
exhaustion.

\section{Properties of pulsations in BSG models}
In this section we discuss the properties and excitations 
of radial and nonradial pulsations in BSG models with rotational 
mixing.  Although these models start with  40\% of  the critical rotation
at the zero-age main-sequence stage, the rotation speeds in the envelopes
in all the supergiant models are very low as discussed in Appendix, so that
we did not include the rotation effects in our pulsation analyses. 
 
\subsection{Radial pulsation}
\begin{figure} %fig3
\epsfig{file=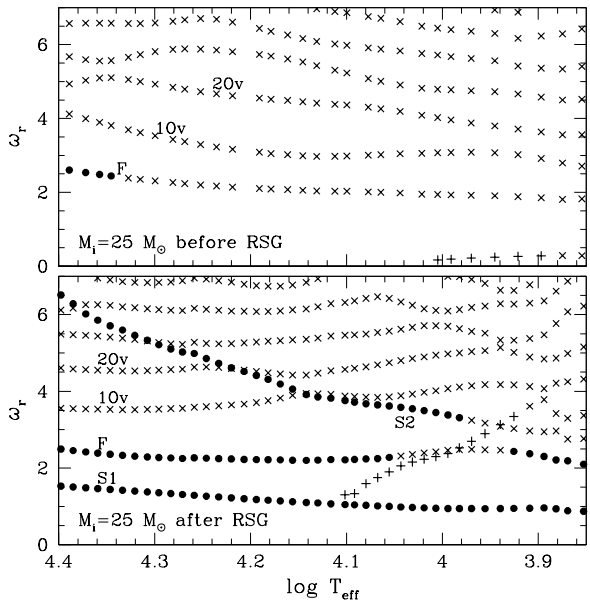,width=0.49\textwidth}
\caption{
Normalized pulsation (angular) frequency, $\omega_{\rm r}$
(the real part of complex eigenfrequency $\omega$), 
for BSG models of $M_i=25M_\odot$ with rotation are plotted as functions of 
$\log T_{\rm eff}$.
The top and the bottom panels are for models evolving toward and from
the RSG stage, respectively.
The masses in the latter models are significantly reduced from the 
former models, while the luminosities are comparable.
Filled circles indicate excited modes, while `$\times$' 
and `+' are damped modes.
The symbol `+' is used for the thermal damping modes with 
$2.0\ge\omega_i/\omega_r\ge0.5$,  for which damping time is
comparable to the pulsation period
(modes with $\omega_i> 2\omega_r$ are not plotted).
}
\label{fig:omega}
\end{figure}

In our linear pulsation analysis, the temporal dependence of variables
is set to be $\exp(i\sigma t)$, where 
$\sigma (= \sigma_{\rm r} +i\sigma_{\rm i})$ is 
a complex frequency obtained as the eigenvalue for the set 
of homogeneous differential equations for linear pulsations. 
The real part $\sigma_{\rm r}$ gives the pulsation period ($2\pi/\sigma_{\rm r}$) 
and the imaginary part $\sigma_i$ gives the stability of the pulsation
mode (excited if $\sigma_{\rm i} < 0$).
We use the symbol $\omega ~(=\omega_{\rm r}+i\omega_{\rm i})$ 
for normalized (complex) frequency; i.e.,
%$$
\[
\omega = \omega_{\rm r}+i\omega_{\rm i} =(\sigma_{\rm r}+i\sigma_{\rm i})/\sqrt{GM/R^3}
\]
%$$ 
with $G$ being the gravitational constant
and $R$ the stellar radius.

Fig.\,\ref{fig:omega} shows $\omega_{\rm r}$, normalized pulsation frequency,  
for low-order radial pulsation modes in the BSG models 
of $M_{\rm i}=25\,{\rm M}_\odot$ 
evolving toward the RSG stage (top panel) and evolving from the RSG 
on a blue-loop (bottom panel).
Filled circles indicate excited modes, while `$\times$' and `+' are damped modes.
In the top panel, the normalized frequency of each mode varies regularly
as a function of $\log T_{\rm eff}$, keeping the order 
such that the lowest frequency
is the fundamental mode (F) with no node in the amplitude distribution, 
next one is the first overtone (1Ov) with one node, and so on.
(The ordering is strictly hold only in adiabatic pulsations).
The fundamental mode in the range $\log T_{\rm eff} \ge 4.34$ is excited by
the Fe-opacity bump as the models are in the $\beta$ Cephei instability region.
The other modes in the top panel are all damped.  
A very low frequency mode which appears in the range $\log T_{\rm eff} \la 4.0$
of the top panel is a mode associated with thermal (damping) wave,
so that it is strongly damped. The symbol '+' is used in Fig.\,\ref{fig:omega}
for strongly damped modes with $2.0\ge\omega_i/\omega_r\ge0.5$ 
(modes with $\omega_i> 2\omega_r$ are not plotted).

In models with high $L/M$ ratios as shown in the bottom panel of 
Fig.\,\ref{fig:omega}, the frequencies of thermal modes 
enter into the frequency range of dynamical pulsations,
and $\omega_i$ decreases (the damping time becomes longer) 
so that the two types of pulsations become indistinguishable. 

In the bottom panel of Fig.\,\ref{fig:omega} for models on the blue-loop, 
at least one mode is excited throughout the  $T_{\rm eff}$ range
(three modes are excited in most part).  
The frequency of each mode varies in a more complex way; 
the mode ordering rule is lost, additional mode sequences appear, and
frequent mode crossings occur, etc. 
The appearance of such complex behaviors is related with strange modes.
The strange modes may be defined as modes which are not seen
in adiabatic analyses. 
With this definition the thermal damping modes also belong to strange modes,
but we are more interested in another type of strange modes which
are excited by a special instability even when the 
thermal time goes to zero which was first recognized by \citet{gau90}.  
(We will discuss the mechanism briefly below).
Sequence S2 in Fig.\,\ref{fig:omega} corresponds to such a strange mode.
Sequence S1, another strange mode, is somehow related with
the lowest frequency mode ($\omega_r<0.05$) seen in the range 
$\log T_{\rm eff} \la 4.0$ of the top panel.
The two sequences is connected in the RSG stage, which is not shown
in Fig.\,\ref{fig:omega}.
The S1 mode, as discussed below, seems to be excited mainly by  
enhanced $\kappa$-mechanism at the Fe-opacity bump. 

The main difference between the models in the top and in the  bottom panels
of Fig.\,\ref{fig:omega} 
is the luminosity to mass ratio;  models in the bottom 
panel (on the blue-loop) have 
$L/M\sim2.0\times10^4\,{\rm L}_\odot/{\rm M}_\odot$, 
while in the top panel   $L/M\sim1.1\times10^4\,{\rm L}_\odot/{\rm M}_\odot$.
A higher $L/M$ ratio makes pulsations more nonadiabatic. 
This can be understood from a linearized form of energy conservation for 
stellar envelope;
\begin{equation}
 T{\partial\delta S\over\partial t} = -{L\over M}{\partial\over\partial q}\left({\delta L\over L}\right),
\label{eq:encons}
\end{equation}
where $\delta$ means the Lagrangian perturbation of the next quantity, $S$ is
the entropy per unit mass, and $q\equiv M_r/M$. 
The above equation indicates that generally a high value of $L/M$ generates
a large entropy change and hence large nonadiabatic effects.

A large $L/M$ ratio has also a significant effect on the envelope structure by
enhancing the importance of radiation pressure. 
Combining a hydrostatic equation with a radiative diffusion equation for the envelope
of a static model, we obtain a relation 
\begin{equation}
{dP_{\rm gas}\over dP} = 
1-{\kappa\over 1.3\times10^4{\rm (cm^2/g)}}{L_{\rm rad}/L_\odot\over M/M_\odot},
\label{eq:pgas}
\end{equation}
where $P_{\rm gas}$ is the gas pressure, $P$ is the total pressure, 
$\kappa$ is the opacity in units of ${\rm cm^2 g^{-1}}$, and $L_{\rm rad}$
is the local radiative luminosity.
This equation indicates that the inward increase
of the gas pressure is hampered or inverted if 
$L_{\rm rad}/M \ga 10^4\,{\rm L}_\odot/{\rm M}_\odot$, 
and the effect is strong where the opacity is large.
A radiation pressure dominated zone  
is formed around an opacity peak in the stellar envelope 
with a $L/M$ ratio larger than $\sim10^4\,{\rm L}_\odot/{\rm M}_\odot$. 
We note that if the second term on the right hand side of 
equation (\ref{eq:pgas}) is sufficiently large, a density inversion is formed.

\begin{figure*} %fig4
\epsfig{file=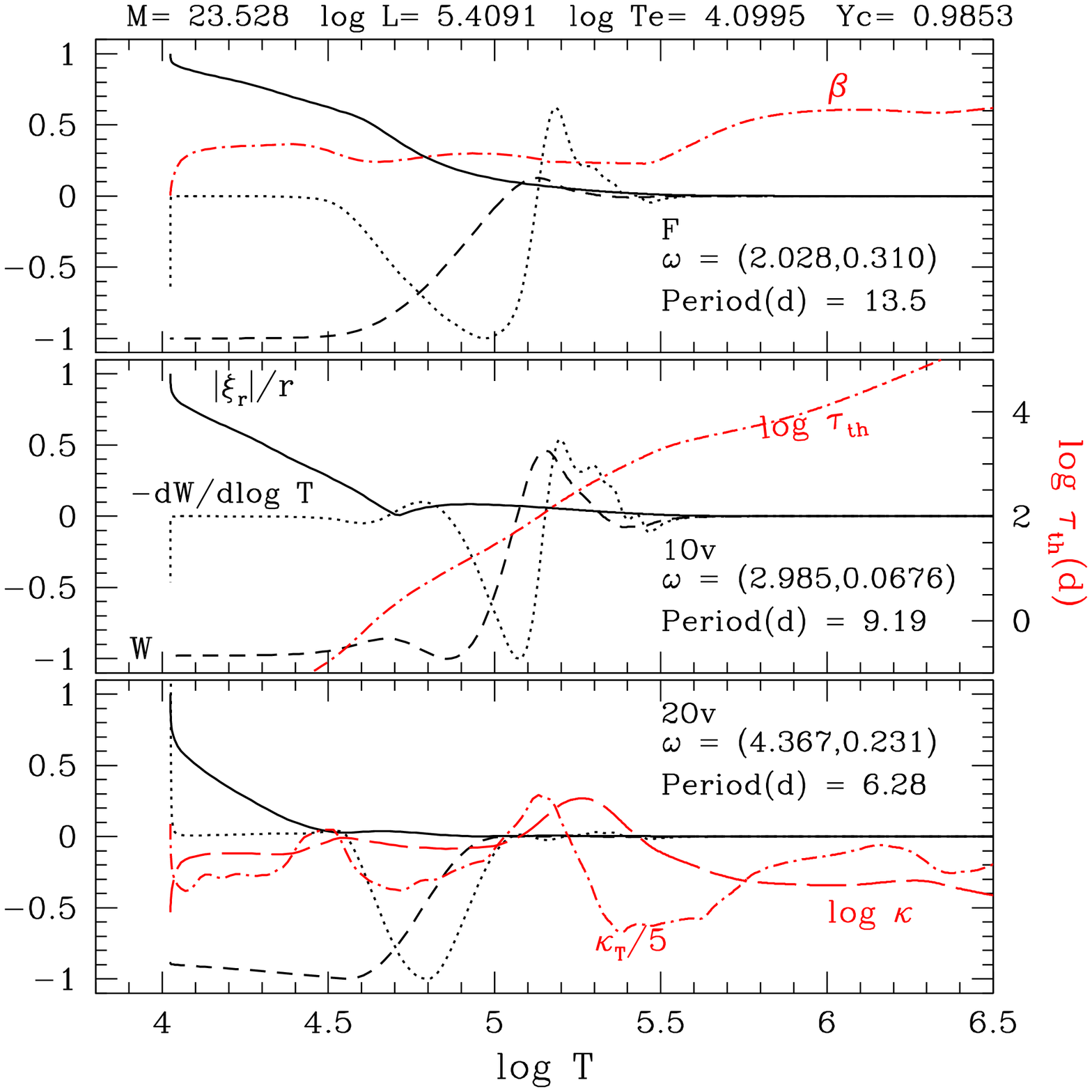,width=0.49\textwidth}
\epsfig{file=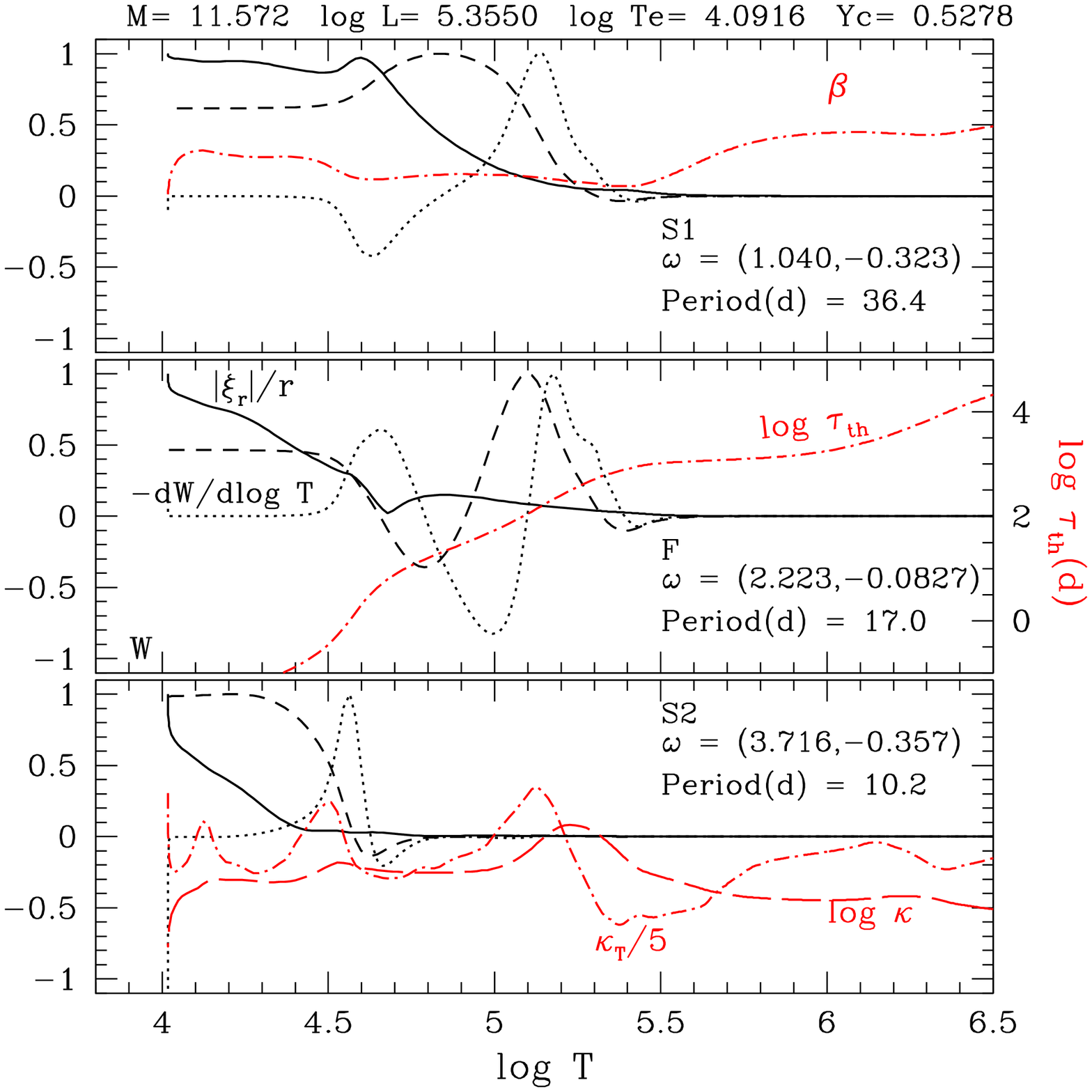,width=0.49\textwidth}
\caption{
Pulsational displacement $\xi_r$, work $W$, and differential work
$-dW/d\log T$ for the three lowest frequency modes are shown
as functions of temperature in BSG models
before (left panel) and after (right panel) the RSG stage.
These quantities are normalized such that the maximum absolute value is unity.
The two models have similar effective temperatures 
of $\log T_{\rm eff}\approx 4.1$.
The mode name given in each panel corresponds to the name given in 
Fig.\,\ref{fig:omega}.
The runs of $\beta =P_{\rm gas}/P$,  $\log \kappa$,
and $\kappa_T =(\partial\ln\kappa/\partial\ln T)_\rho$, and
thermal time $\tau_{\rm th}$ (in days) defined by eq.\,(\ref{eq:tau})  
are also shown. For $\tau_{\rm th}$ the right-side vertical scale of each panel
should be referred.
}
\label{fig:work3}
\end{figure*}

Fig.\,\ref{fig:work3} shows  the properties of 
three pulsation modes F, 1Ov, and 2Ov in a BSG model (with 23.5\,M$_\odot$)
before the RSG stage (left panels)
and S1, F, and S2 modes in a BSG model (with 11.6\,M$_\odot$) 
on the blue-loop after the RSG stage (right panels).
Both models have a similar effective temperature of 
$\log T_{\rm eff}\approx 4.1$. 
Each panel shows the runs of 
fractional displacement, $|\xi_r/r|$ (solid line),  work $W(r)$ (dashed line), 
and differential work $-dW/d\log T$ (dotted line).
The work is defined as 
\begin{equation}
W(r)=4\pi\int_0^rPr^2 {\rm Im}\left({\delta P^*\over P}{\delta\rho\over\rho}\right)dr,
\label{eq:work}
\end{equation}
where $\delta P$ and $\delta\rho$ are the Lagrangian perturbations of the
pressure and the density, respectively, and the superscript $^*$ indicates
the complex conjugate of the quantity.
(In the nonadiabatic linear pulsation analysis, we employ  complex 
representations). 
The layers with $dW/dr \propto -dW/d\log T >0$ (dotted lines) 
contribute to drive the pulsation.
The net effect of driving and damping through the stellar interior
appears in the surface value of the work, $W(R)$; if $W(R) > 0$
the pulsation mode is excited. 
The amplitude growth rate ($-\sigma_i/\sigma_r$) is related with $W(R)$ as 
\begin{equation}
-{\sigma_i\over\sigma_r}={W(R)\over 2\sigma_r^2}
\left[\int_M \bm{\xi}^*\cdot\bm{\xi}dM_r\right]^{-1},
\end{equation}
where $\bm{\xi}$ is pulsational displacement vector (for radial pulsations
$\bm{\xi}=\xi_r\bm{e}_r$ with $\bm{e}_r$ being the unit vector 
in the radial direction). 
We assume the amplitude of an excited radial-pulsation mode 
to grow to be visible.
All modes in the left panels of Fig.\,\ref{fig:work3} are damped,
while all modes in the right panels are excited.

Some structure variables are also shown in Fig.\,\ref{fig:work3}.
Note that in the model on the blue-loop (right panel), 
$\beta (= P_{\rm gas}/P)$ is 
extremely small in the range $4.6 \la \log T \la 5.4$, indicating 
$P\approx P_{\rm rad}$ there. 
The thermal time $\tau_{\rm th}$ is defined as
\begin{equation}
\tau_{\rm th}(r) = \int_r^R4\pi r^2 {\rho C_p T\over L_r} dr,
\label{eq:tau}
\end{equation}
where $C_p$ is the specific heat per unit mass at constant pressure.
In the outermost layers with $\log T\la 4.7$, $\tau_{\rm th}$ is shorter
than pulsation periods.
The pulsations are locally very non-adiabatic there.

For the longest period modes shown in the top panels in Fig.\,\ref{fig:work3},
driving occurs around the Fe-opacity bump ($\log T\sim 5.0 - 5.4$).
Since the thermal time there is longer than the pulsation periods,
the driving can be considered as the ordinary $\kappa$-mechanism.
Roughly speaking, the $\kappa$-mechanism works (under a weak nonadiabatic
environment) if the opacity derivative with respect to temperature increases 
outward; i.e., $d\kappa_T/dr >0$ \citep{unn89}.
We see that this rule is hold in the model before the RSG stage (left panel),
in which the driving is overcome by radiative damping in the upper layers
(where $d\kappa_T/dr <0$) so that the mode is damped.
For the longest period mode (S1) in the model on the blue-loop (right panel), 
the driving zone extends out into zones where $\kappa_T$ decreases outward.
Because of the extension of the driving zone, 
which is probably caused by small $\beta$,
the mode is excited; i.e., the driving effect exceeds radiative damping
in the upper layers. 
This mode is considered to be a strange mode because there is 
no adiabatic counterpart. However, the excitation is caused by the enhanced 
$\kappa$-mechanism, in accordance with the finding of \citet{dzi05}. 
  
The modes in the middle panels in Fig.\,\ref{fig:work3} are ordinary modes;
the first overtone, 1Ov, for the model before the RSG stage (left panel) 
and fundamental mode, F,  for the model on the blue-loop (right panel).
For these modes the driving around $\log T \sim  4.5 - 4.8$ has some
contribution in addition to the driving around the Fe-opacity bump.
The mode in the left panel 
is damped because radiative damping
between the two driving zones exceeds the driving effects, while 
the F mode in the right panel is excited. 
The driving in the low temperature zone should not be 
the pure $\kappa$-mechanism
because the thermal time there is shorter than the pulsation periods.
It should be somewhat affected by the strange mode instability, 
which is discussed below.

The mode in the left-bottom panel in Fig.\,\ref{fig:work3}, the second
overtone (2Ov) of the model before the RSG stage 
is damped because of the lack of appreciable drivings.
On the other hand, the S2 mode in the right-bottom panel 
is excited strongly in the zone ranging 
$4.3 \la \log T \la 4.6$ (HeII ionization zone), 
where the thermal time $\tau_{\rm th} < 1$\,d is much shorter than the
pulsation period (10.2\,d).
Because no heat blocking (which is essential for the $\kappa$-mechanism) 
occurs there,
the driving mechanism must be the genuine strange-mode instability, which
should work even in the limit of $\tau_{\rm th} \rightarrow 0$; 
NAR (Nonadiabatic reversible) approximation introduced by \citet{gau90}.
In this limit, $\delta L = 0$ (cf. eq.\,(\ref{eq:encons})). 
From this relation with the plane parallel approximation 
it is possible to derive an approximate relation of 
\begin{equation}
\delta P \propto \pm i\kappa_\rho \kappa F_{\rm rad}{\delta\rho\over\rho}
\end{equation}
\citep{sai09}, where $F_{\rm rad}$ is the radiative flux and
$\kappa_\rho\equiv (\partial\ln\kappa/\partial\ln\rho)_T$.
This relation indicates that a large phase difference arises between
$\delta P$ and $\delta \rho$, which
can lead to strong driving (and damping)
according to the work integral given in eq.\,(\ref{eq:work}); 
in the limit of NAR
approximation, if $\sigma$ is an eigenvalue, 
the complex-conjugate $\sigma^*$ is also an eigenvalue.  
This explains the strange mode instability 
\citep[see][for a different approach]{gla94}.

\subsection{Nonradial pulsations}

\begin{figure*} %fig5
\epsfig{file=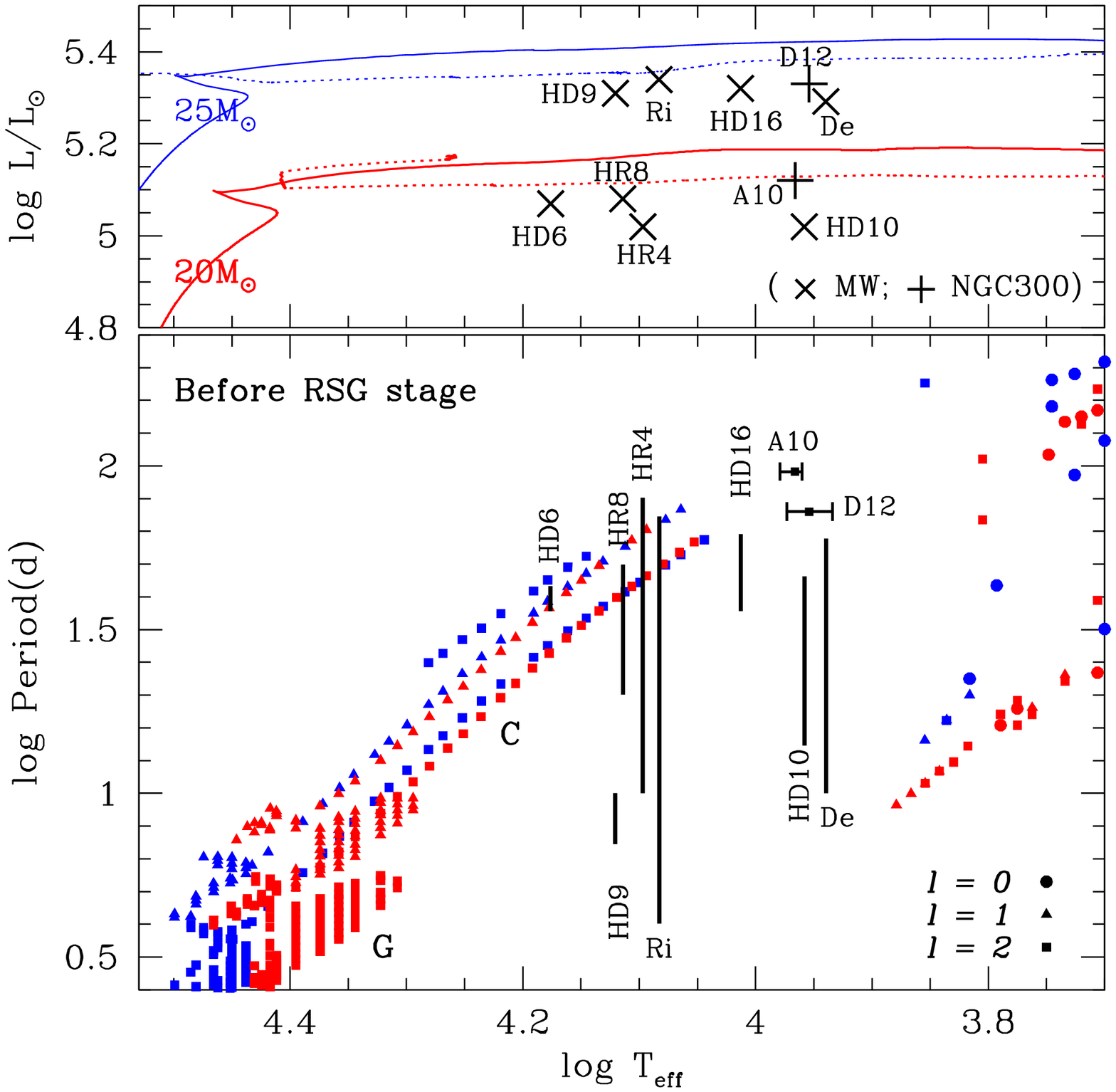,width=0.49\textwidth}
\epsfig{file=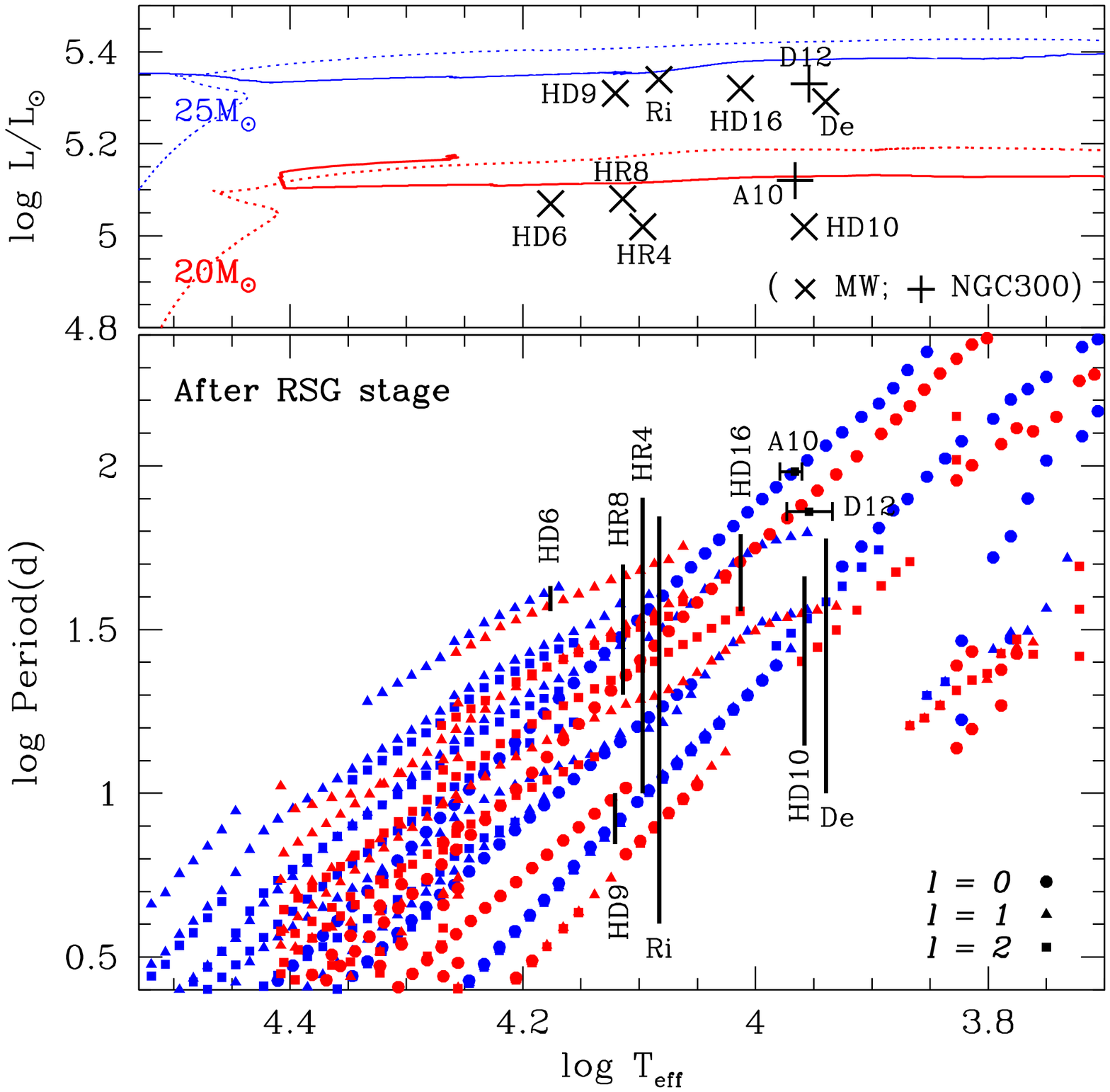,width=0.49\textwidth}
\caption{
Evolutionary tracks of $M_{\rm i}=20$ and 25\,M$_\odot$ models 
with rotation (top panels)
and periods  of pulsations excited (and visible satisfying eq.(\ref{eq:ratio}))
in the models (bottom panels) 
are compared with some of the $\alpha$ Cygni variables
in our  Galaxy and NGC\,300.
The left panels are for the models evolving toward the RSG stage,
while the right panels are for the models on the blue loop from the RSG stage.
The periods (or period ranges) are shown for some $\alpha$ Cygni variables 
in the Milky Way and in NGC\,300 having luminosities 
within the considered range.
Adopted parameters and periods are listed in Table\,\ref{tab:param}.
Names of stars;
De=Deneb ($\alpha$ Cyg), HD10=HD100262, HD16=HD168607,
Ri=Rigel ($\beta$ Ori), HR4=HD96919 (HR4338), HR8=HD199478 (HR8020),
HD9=HD91619, and HD6=HD 62150; 
and A10 and D12 are star names in NGC\,300 used by \citet{bre04}.
}
\label{fig:teperi}
\end{figure*}

The three dimensional property of a nonradial pulsation of a spherical star
is characterized by the degree $l$ and the azimuthal order $m$ 
of a spherical harmonic $Y_l^m$
($l=1$ for dipole and $l=2$ for quadrupole modes). \footnote{The pulsation
frequency does not depend on $m$ in a non-rotating and non-magnetic
spherical star.}
There are two types of nonradial pulsations; 
p-modes (common to radial modes) and 
g-mode pulsations. The g-mode pulsations are possible only in the 
frequency range below the Brunt-V\"ais\"al\"a 
(or buoyancy) frequency \citep[e.g.,][]{unn89,aer10}. 
We have performed nonradial pulsation analyses based on the method
described in \citet{sai80}, disregarding the effect of rotation.
This is justified because we discuss the modes trapped in the envelope, 
where the rotation speed is very low as discussed in Appendix.

The properties and the stability of nonradial pulsations of supergiants are 
very complex because they have a dense core with very high
Brunt-V\"ais\"al\"a frequency.
All oscillations in the envelope with frequencies less than the maximum
Brunt-V\"ais\"al\"a frequency in the core can couple with g-mode oscillations
in the core through the evanescent zone(s) laying between the envelope 
and the core cavities.
The coupling strength varies sensitively with the pulsation frequency and 
the interior structure of the star.
Depending on the coupling strength, the relative amplitudes in the core and in
the envelope vary significantly, and hence the stability changes.

Excitation by the $\kappa$-mechanism and the strange-mode instability works
also for nonradial pulsations \citep{gla96}.
In addition, oscillatory convection mechanism \citep{shi81,sai11} 
works in the convection zone associated 
with the Fe-opacity bump in the BSGs. 
Furthermore, the $\epsilon$-mechanism of excitation at the H-burning shell 
can excite nonradial modes which are strongly confined to a narrow
zone there as shown by \citet{mor12b}.

Among those nonradial modes which are excited,  
we restrict ourselves, in this paper, 
to possibly observable modes having the following properties:
%\[
\begin{equation}
l \le 2  \quad {\rm and} \quad  
f_{\rm amp}\equiv{|\xi_r|_{\rm surf}/R\over (|\xi_r|/r)_{\rm max}} > 0.1,
\label{eq:ratio}
\end{equation}
%\]
where $\xi_r$ is the radial component of Lagrangian displacement,
and the subscripts $_{\rm surf}$ and $_{\rm max}$ indicate the values at 
the stellar surface and at the maximum amplitude in the interior, respectively. 
The first requirement selects modes less affected by 
cancellation on the stellar surface. 
The second requirement excludes modes highly trapped in the interior; i.e., 
such oscillations hardly emerge to the surface.
The second requirement excludes most of the g-modes excited in the core,
because they are strongly trapped in the core having small ratios of 
$f_{\rm amp} < 10^{-3}$.

Fig.\,\ref{fig:teperi} shows periods of excited nonradial dipole and quadrupole
modes ($l=1, 2$) (as well as radial modes) in models evolving 
toward RSG region (left panel)
and in models on the blue-loop after the RSG stage (right panel) 
for $M_{\rm i} = 20$ and  25\,M$_\odot$
cases with rotation. 
Obviously, much more modes are excited in the BSG models  after the 
RSG stage (on the blue-loop) in the period ranges of $\alpha$ Cygni variables.

In the BSG models before the RSG stage, excited observable modes are
nonradial g-modes and oscillatory convection modes.
Swarms of modes labeled as 'G' in the left panel of Fig.\,\ref{fig:teperi}
are g-modes excited by the Fe-opacity bump.
For those oscillations, the amplitude is confined to the envelope by the
presence of a shell convection zone above  the H-burning shell 
\citep{sai06,god09,gau09},
which prevents the oscillation from penetrating into and being 
damped in the dense core. 
The red edge for the group, $\log T_{\rm eff}\approx 4.3$,  is bluer by 0.1 dex 
than that obtained by \citet{gau09} for 25\,M$_\odot$ models with $Z=0.02$.
The difference can be attributed to the metallicity difference.
These g-modes are probably responsible for the multi-periodic variations of the
early BSG HD 163899 (B2Ib/II) \citep{sai06} observed by 
the MOST satellite and some of the 
relatively less luminous early BSGs studied by \citet{lef07}.

The sequences labeled as `C' in the left panel of Fig.\,\ref{fig:teperi}
are oscillatory convection (g$^{-}$) modes associated
with the convection zone caused by the Fe-opacity peak around $\log T\sim 5.2$
\citep[the presence of such modes is discussed in][]{sai11}.
The sequences terminate when the requirement of 
$f_{\rm amp} > 0.1$ is not met anymore; 
i.e., beyond the termination the modes are trapped
strongly in the convection  zone.

\begin{figure} %fig6
\epsfig{file=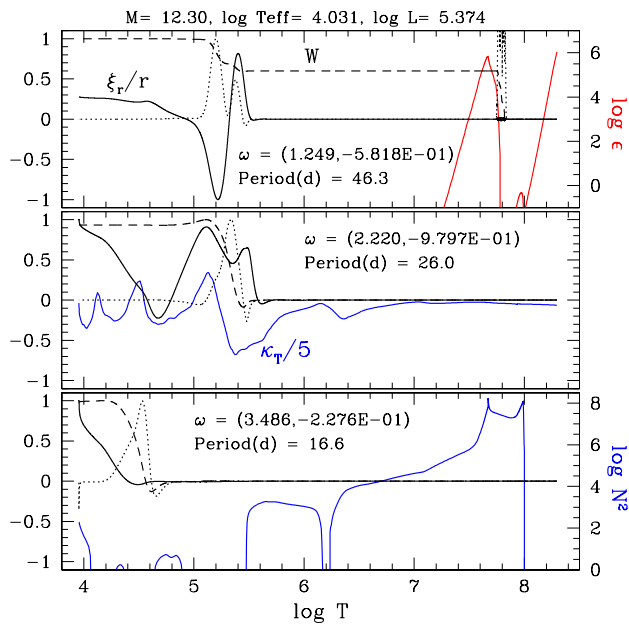,width=0.49\textwidth}
\caption{Three nonradial dipole ($l=1$) modes excited in a BSG model
($M_{\rm i}=25M_\odot$) after the RSG stage.
Black solid lines show the real part of radial displacement, $\xi_r/r$.
Dashed lines are for the work, $W(r)$ defined by eq.\,(\ref{eq:work}).
Dotted lines are for the differential work $-dW/d\log T$.
These quantities are plotted as functions of temperature in the model;
they are normalized such that the maximum absolute value in the 
envelope is unity.
The blue lines in the bottom and the middle panel show the run of square 
of Blunt-V\"ais\"al\"a frequency, $N^2$ (normalized by $GMR^{-3}$)
and the opacity derivative $\kappa_T$, respectively. 
The red line in the top panel shows the run of nuclear
energy generation rates $\epsilon$(erg\,g$^{-1}$s$^{-1}$).
}
\label{fig:nradwork}
\end{figure}

The right panel for models
after the RSG stage show many modes excited in the $T_{\rm eff}$--period
range appropriate for the $\alpha$ Cygni variables.
They are excited by different ways depending on the periods.
Fig.\,\ref{fig:nradwork} presents examples, 
in which amplitude and work curves are shown for the three dipole ($l=1$)
modes excited (and $f_{\rm amp}>0.1$; eq.(\ref{eq:ratio})) in a model
of $M_{\rm i}=25\,{\rm M}_\odot$ at $\log T_{\rm eff} = 4.031$ 
and $\log L/{\rm L}_\odot=5.374$.
(Note that the mass of the model is reduced to 12.3\,M$_\odot$ and hence
the $L/M$ ratio is as high as $\sim2\times10^4\,{\rm L}_\odot/{\rm M}_\odot$.)

The top panel of Fig.\,\ref{fig:nradwork} shows the lowest frequency 
dipole mode excited.
This is one of the very rare cases in which the driving effect in the core 
 is comparable with that in the envelope.
 In the envelope the amplitude is strongly confined to 
 the Fe convection zone, which is characteristic of oscillatory 
convection modes.
The envelope mode weakly couples with a core g-mode.
Although the amplitude in the core is extremely small, the driving effect is
comparable or larger than that in the envelope.
In the hotter models contribution from the core is negligibly small, while
as $T_{\rm eff}$ decreases the core contribution increases rapidly but 
the ratio $f_{\rm amp}$ soon becomes much smaller than 0.1.

The modes shown in the middle and the bottom panels of Fig.\,\ref{fig:nradwork}
are confined strongly to the envelope without any contribution from the core.
The mode in the middle panel with a period of 26 days is 
excited around Fe-opacity 
bump at $T\sim 2\times10^5$K and have large amplitude in the convection zone.
The mode in the bottom panel with a period of 16.6\,days
corresponds to the S2 strange mode of radial pulsations shown 
in Fig.\,\ref{fig:work3} (right-bottom panel),  excited by the strange-mode
instability around the He II ionization.

Three quadrupole ($l=2$) modes having periods and properties similar
to those of dipole modes shown in Fig.\,\ref{fig:nradwork}
are also excited in the same model. 
However, the amplitudes of the two longer-period quadrupole modes
are more strongly confined in the Fe-convection zone with
amplitude ratios of $f_{\rm amp} \sim 0.05$ so that they do not meet the 
requirement given in eq.\,(\ref{eq:ratio}) and hence are not considered to be  
observable.
Only the shortest period (16.1 days) mode with $f_{\rm amp} =1$ 
satisfies the requirement; this mode corresponds to the S2 mode. 

In this model two radial strange modes belonging to S1 and S2 are
also excited (Fig.\,\ref{fig:omega}) 
with periods of 54.1 and 16.3\,days, respectively.
Counting these excited observable pulsation modes,
we expect a total of six pulsation periods;  
54.1 and 16.3\,days for the two radial pulsations,
46.3, 26.0, and 16.6\,days for the three dipole modes, 
and 16.1\,days for the quadrupole mode.
The three very close periods of $\sim 16$\,days would 
yield very long beat periods up to a few years. 
If these modes are simultaneously excited, the light curve would be
extremely complex, which is just consistent with light curves of many 
$\alpha$ Cygni variables \citep[e.g.,][]{van98}.

\section{Discussions}
\subsection{Comparison with periods of $\alpha$ Cygni variables}
Despite the long history of observations for $\alpha$ Cygni variables, periods of
variations are only poorly determined for most of the cases, hampered by
complex and long-timescale light and velocity variations.
Here we compare observed periods of relatively less luminous $\alpha$ Cygni
variables shown in Fig.\,\ref{fig:teperi} with theoretical models
of $M_{\rm i}=20, 25\,M_\odot$ with rotational mixings.

\begin{table}
\caption{Adopted parameters and periods of $\alpha$ Cygni variables 
in the Galaxy and NGC\,300} 
\label{tab:param}
\begin{tabular}{@{}lcccccc@{}}
\hline
HD  &  $\log T_{\rm eff}$  & $\log L$ 
& ref & Peri.(d) & ref\cr
\hline
\multicolumn{6}{c}{MW Galaxy}\cr
34085 (Rigel) &  4.083  & 5.34 & a &    4--70  & b \cr
62150 & 4.176  & 5.07 & c & 36--43 & d \cr
91619 %(V369 Car)   
& 4.121 & 5.31 &  e,f &   7--10 & g \cr
96919 (HR4338) 
& 4.097 & 5.02 & f &     10--80    & g \cr           
100262 %(omi02 Cen) 
&  3.958 &  5.02 & h  &   14--46 & g,i\cr  
168607 %(V4029 Sgr) 
&  4.013 & 5.32 & c &  36--62 & d,j \cr  
197345 (Deneb)  & 3.931  & 5.29 & k &  10--60 & l \cr
199478 (HR8020) & 4.114 & 5.08 & m &   20--50 & n \cr 
\multicolumn{6}{c}{NGC\,300}\cr
A10 & 3.966$^{+0.013}_{-0.006}$ & 5.12 & o & 96.1 & p \cr
D12 & 3.954$^{+0.019}_{-0.020}$& 5.33 & o & 72.5 & p \cr
\hline
\end{tabular}\\
a =\citet{fir12};~ %1
b=\citet{mor12a};~%2
c=\citet{lei84};~%3
d=\citet{van98};~%4
e=\citet{fra10};~%5
f=\citet{kal10};~%6
g=\citet{kau97};~%7
h=\citet{kau96};~%8
i=\citet{ste77};~%9
j=\citet{ste99};~%10
k=\citet{sch08};~%11
l=\citet{ric11};~%12
m=\citet{mar08};~%13
n=\citet{per08};~%14
o=\citet{kud08};~%15
p=\citet{bre04}   %16
\end{table}

\subsubsection{Deneb ($\alpha$ Cyg)}
Deneb is the prototype of the $\alpha$ Cygni variables and has a long history
of studies \citep[see][for a review and references given]{ric11}.
Although its light-curves are very complex, 
\citet{ric11} obtained radial pulsation features at two short epochs. 
Strangely, however, 
the two epochs give two different periods; 17.8\,days and 13.4\,days.
From the time-series analyses by \citet{ric11} and \citet{kau96},
we have adopted a period range of $10-60$\,days.
On the HR diagram Deneb is close to the blue-loop of 
$M_{\rm i}=25\,{\rm M}_\odot$
(Fig.\,\ref{fig:teperi}); the current mass is about 12.7\,M$_\odot$. 
Our model at $\log T_{\rm eff}\approx3.93$ predicts the following periods 
for the excited modes; 127 \& 49\,days (radial), 42\,days ($l=2$).
The longest period is longer than the observed period range.
The other two periods are in the longer part of the observed range.
The periods of excited radial modes are much longer than the periods
of radial pulsations, 17.8 and 13.4\,days,  obtained by \citet{ric11}.
The reason for the discrepancy is not clear.
Our model cannot explain the shorter part of the observed period range.

\citet{gau92} tried to explain Deneb's pulsations by nonradial strange modes and
found that if the mass is less than 6.5\,M$_\odot$, nonradial modes with various
$l$ are excited.  The required mass seems too small for the position of Deneb
on the HR diagram.  
Recently, \citet{gau09} found that  nonradial
modes of $l=1,2$ having periods consistent with the observed period range 
are excited by the H-ionization zone ($T\sim10^4$\,K) 
in models of 25\,M$_\odot$ 
(evolving towards the RSG stage)
with effective temperatures similar to the observed one.
In our models, however, the H-ionization zone excites relatively short 
periods modes in cooler models with $\log T_{\rm eff} \sim 3.8 - 3.85$
(Fig.\,\ref{fig:teperi}), which is inconsistent with Deneb.  
The reason for the difference is not clear; 
further investigations are needed.

\subsubsection{Rigel ($\beta$ Ori)}
The period range given in Table\,\ref{tab:param}, $4-70$\,days, is based on the 
recent analysis by \citet{mor12a}. 
\citet{kau97}' s analysis indicates a narrower range of $5-30$\,days.
The position on the HR diagram (Fig.\,\ref{fig:teperi}) is consistent with a 
$M_{\rm i}=25\,{\rm M}_\odot$ model on the blue-loop after the RSG stage.
Our model at $\log T_{\rm eff}\approx 4.08$ predicts excitation of the following 
visible modes:
40.1, 18.4, and 11.2\,days (radial); 31.9, 16.4, and 11.4\,days ($l=1$); 
11.0\,days ($l=2$). 
Our model predicts many pulsation modes in the longer part of the observed
period range.
Although no modes with periods shorter than 10 days are excited,
those shorter periodicities might be explained by combination frequencies
among excited modes.
\citet{mor12b} proposed g-modes (with periods of longer than 20\,days) 
excited by the $\epsilon$-mechanism at
the H-burning shell. However, we think that those modes are invisible
because the amplitude is very strongly confined to a narrow zone close to the
H-burning shell.

\subsubsection{HD 62150}
The period range of HD\,62150, $36-43$\,days, 
is adopted from the analysis by \citet{van98} 
of the Hipparcos observations. In fact they obtained a 36.4 day period and a
very small amplitude period of 43.0 days.
These periods are consistent with nonradial (oscillatory convection
mode) pulsations either in the first crossing toward the RSG stage,
 or on the blue loop of
a model with $M_{\rm i} \approx20M_\odot$ (Fig.\,\ref{fig:teperi}). 
HD 62150 is, among the variables considered here,  
the only star which can be considered as in the first crossing stage. 

\subsubsection{Other stars in the MW}
HD100262 has a similar $T_{\rm eff}$ to Deneb and has the same problem; 
i.e., the model cannot predict the shorter part of the observed period range.
The period ranges of other stars, HD 168607, HR4338, HR 8020, and HD 91619 
are roughly consistent with predicted periods of the models evolving 
on blue-loops from the RSG region.

\subsubsection{A10 and D12 in NGC\,300}
\citet{bre04} found many $\alpha$-Cygni type variables among 
blue supergiants in the galaxy NGC\,300.
Among them, two stars, A10 and D12 show regular light curves 
with periods of 96.1\,d and 72.5\,d, respectively. 
The light curves look consistent with radial pulsations. Hence, according
to our stability results, these stars must be on blue-loops after the RSG stage.
\cite{kud08} obtained heavy element abundances of 
$\log Z/Z_\odot = -0.04\pm0.20$
for A10 and $-0.16\pm0.15$ for D12, which indicate our models with $Z=0.014$
to be appropriate for these stars.
The periods of A10 and D12 roughly agree with theoretical periods 
of radial pulsations obtained
for  $M_{\rm i}=20$ and 25\,M$_\odot$ models (Fig.\,\ref{fig:teperi}, right panel).
These are strange modes excited at the Fe-opacity bump (Fig.\,\ref{fig:work3}) 
in BSG models after RSG stages.
Because of the wind mass loss occurred in the previous stages, 
these models currently have masses of 8.8 and 12.7\,M$_\odot$, respectively.  
Our models agree with  \citet{dzi05} who concluded that the pulsations of
A10 and D12 should be strange modes in models with masses reduced significantly by mass loss.

However, Fig.\,\ref{fig:teperi} (right panel) indicates a discrepancy;
in the HR diagram A10 and D12 are close to the evolutionary 
tracks of $M_{\rm i}=20$
($\log L/{\rm L}_\odot \approx 5.1$) and $25\,{\rm M}_\odot$ 
($\log L/{\rm L}_\odot \approx 5.3$),
 respectively, while the periods correspond to the other way 
around; i.e., the period of A10\,(96.1\,d) is longer than that of D12\,(72.5\,d).
According to the spectroscopic analysis by \cite{kud08}, both stars have
similar effective temperatures, but the radius of A10 (142\,R$_\odot$) is  smaller
than that of D12 (191\,R$_\odot$) due to the luminosity difference; 
i.e., the smaller star has the longer period.
The discrepancy would be resolved if A10 were cooler 
and  D12 were hotter slightly beyond  the error-bars.
In addition, the folded light-curve of A10 shown in Fig.\,3 in \citet{bre04} has
 considerable scatters, indicating other periods might be involved.
Further observations and analyses for the two important stars are 
desirable.

\subsubsection{HD 50064}
\citet{aer10b} found a 57 day periodicity in HD 50064 from the CoRoT
photometry data and interpreted the period as a radial strange mode
pulsation. From their spectroscopic analysis they estimated
$T_{\rm eff} \sim 13500$K ($\log T_{\rm eff} \sim 4.13$), 
$\log g \sim 1.5$, $R\approx200\,{\rm R}_\odot$, 
and $\log L/{\rm L}_\odot \approx6.1$. 
The star seem more luminous than our models.
From these parameters we can estimate 
the normalized frequency corresponding to the observed period, 
using the relation $\omega_{\rm r} = (2\pi/\Pi)\sqrt{R/g}$ with $\Pi$ 
being pulsation period.
Substituting above parameters into the equation 
we obtain $\omega_r\approx 1.3$.
This value should be considered consistent with either S1 or S2 strange mode
(Fig.\,\ref{fig:omega}), taking into account the possibility of 
a considerable uncertainty in the value of $R/g$.
This confirms \citet{aer10b}'s interpretation of the 57 day period 
as a radial strange mode pulsation.

\subsection{Surface CNO abundances}

\begin{table}
\caption{The surface He abundance ($Y$) and ratios of CNO elements 
for BSG models at $\log T_{\rm eff}=4.0$ }
\begin{tabular}{@{}lccc|ccc@{}}
%\begin{tabular}{@{}lcccccc@{}}
\hline
&\multicolumn{3}{c}{BSG before RSG} & \multicolumn{3}{c}{BSG after RSG} \cr
\hline 
$M_{\rm i}$ & $Y$ & N/C & N/O& $Y$ & N/C & N/O \cr
14 (rot) & 0.29 & 2.27 & 0.517 & 0.55 & 38.0 & 2.41\cr
20 (rot) & 0.31 & 2.46 & 0.609 & 0.57 & 39.7 & 2.94 \cr
25 (rot) & 0.35 & 3.23 & 0.877 & 0.64 & 60.4 & 4.22  \cr
\hline
\end{tabular}\\
The initial values are N/C$\equiv X_{\rm N}/X_{\rm C}=0.289$ and
N/O$\equiv X_{\rm N}/X_{\rm O}=0.115$, 
where $X_{\rm i}$ means mass fraction of element i.
The initial helium mass fraction is 0.266.
\label{tab:cno}
\end{table}

As we discussed above, observed periods of many $\alpha$ Cygni
variables are consistent with radial and nonradial pulsations
excited in BSG models evolved from the RSG stage. Because
of the convective dredge-up in the RSG stage, rotational
mixing, and wind mass loss, the surface compositions of the
CNO elements are significantly modified from the original
ones.  

\begin{figure} %fig7
\epsfig{file=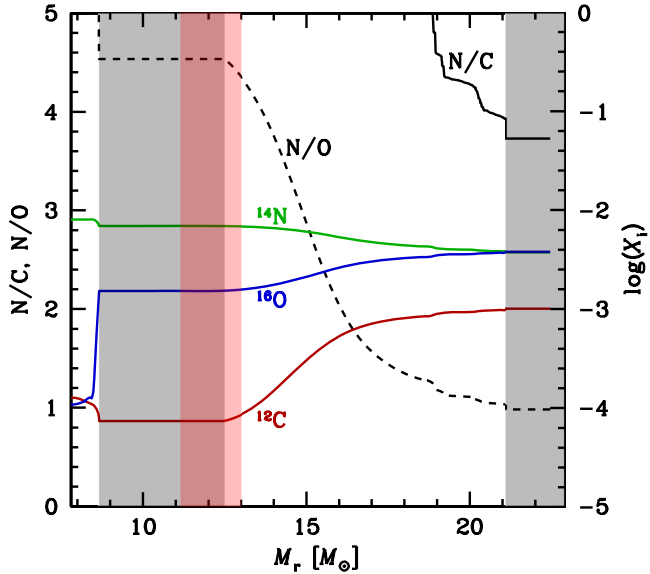,width=0.48\textwidth}
\caption{
Profile of the CNO abundances through the stellar envelope, as well as the profile of N/C and N/O ratios for the rotating $25\,M_\odot$ when it reaches for the first time the RSG branch ($\log(T_{\rm eff})\la 3.6$). The x-axis is the Lagrangian mass coordinate, and we show only the region above the convective core ($7.80\,M_\odot$). The light grey zones correspond to convective layers. The light red zone corresponds to the region with $0.6 < M_{\rm core}/M_r < 0.7$, corresponding to minimum values of the ratio of the core mass to the total mass that is required to have a blue loop.
}
\label{fig:cno}
\end{figure}

To understand the evolution of the surface composition, we show on 
Fig.~\ref{fig:cno} a profile of the CNO abundance through the stellar 
envelope, as well as the profile of N/C and N/O ratios for the 
rotating $25\,M_\odot$ model when it reaches for the first time the red 
supergiant branch ($\log(T_{\rm eff}$) around 3.6, 
as a function of the Lagrangian mass coordinate. 
The edge of the convective core is on the left (Lagrangian mass coordinate 
$7.80\,M_\odot$), and the stellar surface on the right (Lagrangian mass 
coordinate $22.9\,M_\odot$).

The convective zones are shown in transparent gray. The first one 
(between $\sim 9\,M_\odot$ and $\sim 13\,M_\odot$)
 is the convective zone developing on top of the hydrogen shell burning zone. 
The second one is the convective zone below the surface. 
We see that the region in the first convective zone and above 
(up to $\sim 17\,M_\odot$) is strongly affected by CNO-cycle burning 
products, with a large $^{14}$N mass fraction, and a depletion of 
$^{12}$C and $^{16}$O. In this region, both N/C and N/O ratios are very large. 
Only the region near the surface exhibits C, N, and O abundances close 
to the initial ones, with N/C and N/O ratios less than 1.

In Fig.~\ref{fig:cno}, the red region shows the typical ratios 
$M_{\rm core}/M_r$ needed for a star to evolve again towards the blue 
\citep{gia67}. This means that the star will become bluer again only 
if it loses its mass up to that region. The Lagrangian mass coordinate 
corresponding to $M_{\rm core}/M_r = 0.6$ is $\sim13\,M_\odot$, 
so it means that for the core to contain 
60\% of the mass, the star must lose about $9.9\,M_\odot$.
 We can thus expect that a blue star that has evolved from the RSG stage 
should exhibit very high N/O and N/C ratios. 
 {\it The figure shows that what increases a lot the N/C and N/O ratios is not so much the dredge up process due to convection but the mass loss.}

Table\,\ref{tab:cno} shows surface helium abundance and ratios of CNO 
elements (mass fraction) for BSG models (with rotational mixing) 
at $\log T_{\rm eff}\approx4.0$ before and after the RSG stage.
Most of the stars listed in Table\,\ref{tab:param} do not have 
measured N/C and N/O ratios and thus cannot be used 
to test the above predictions. 
There are however two stars Rigel and Deneb for which such data exist. 
\citet{prz10} %Przybilla et al. (2010) 
lists (Y, N/C,N/O)= (0.32, 3.4, 0.65) for Deneb and (0.37, 2.0,0.46) for Rigel. 
These numbers are rather consistent with
models before the RSG stage, in contradiction with our conclusion 
that they should rather be stars having evolved through a RSG stage. 

So we are left here with a puzzle: pulsation properties tell us that 
these two stars should be BSG evolved from a RSG stage, while surface 
abundances indicate that they should be BSG having directly evolved from 
the MS phase.
In other words, in order to have 
the excited modes compatible with the observations of Deneb and Rigel, 
a high $L/M$ ratio is needed, which, in turn, implies 
strongly changed N/C and N/O ratios
which are not observed.

So we are left with a real puzzle here. 
At this point we can simply mention three directions which may help 
in resolving this discrepancy:

1) Are we sure that all the frequency measured correspond to pulsation process?
For instance,  some variations could be due  
to some other type of instabilities at the surface. 
In this case, the vertical lines associated to Rigel and Deneb 
in Fig.\,\ref{fig:teperi} 
representing the observed pulsation period ranges 
might be reduced and be compatible with stars
on their first crossing.

2) It would be extremely interesting to obtain accurate measurements of 
the N/C and N/O ratios at the surface of the stars listed 
in Table\,\ref{tab:param} 
other than Rigel and Deneb to see whether 
the case of Deneb and Rigel are representative of all these stars or not.
In particular, measurements of surface CNO abundances of 
the two radial pulsators A10 and D12
in NGC\,300  are most interesting.

3) 
Recently some authors \citep{dav13} %(Davies et al., submitted to ApJ ) 
find that
the RSG have significantly higher effective temperatures and 
are hence more compact for a given luminosity. 
From another perspective, \citet{des11}  find that to reproduce 
the light curve of type II-P supernovae, the RSG progenitor 
should be more compact than predicted by current models. 
The effective temperature of the RSG stars depends on the physics 
of the convective envelope. 
For instance, depending whether turbulent pressure is accounted for 
or not and how the mixing length is computed, much bluer positions 
for the RSG stars can be obtained
\citep[see e.g. Fig. 9 in][]{mae87}. 
%Maeder and Meynet 1987, A\&A, 182, 243). 
One can wonder whether more compact RSG stars
would need to lose as much mass as larger RSG stars to 
evolve to the blue part of the HR diagram.
Would the blue supergiants resulting from the evolution of 
more compact RSG stars present the
same chemical enrichments as those presented by the current models?
We shall investigate these questions in a forthcoming work.

\section{Conclusions}
We have studied the pulsation properties of  BSG models.
In the effective temperature range between the blue edge of the 
cepheid instability
strip and the red-edge of the $\beta$ Cephei instability range, where many
$\alpha$ Cygni variables reside, radial (and most of nonradial) pulsations
are excited only in the models evolving on a blue-loop 
after losing significant mass in the RSG stage. 
The observed quasi periods of $\alpha$ Cygni variables are found to be roughly
consistent with periods predicted from these models in most cases.
This indicates that the $\alpha$ Cygni variables are mainly He-burning 
stars on the blue-loop. 

However, it is found that the abundance ratios N/C and N/O on the surface
seem too high compared with spectroscopic results.
Further spectroscopic and photometric investigations for BSGs
are needed as well as theoretical searches for missing physics in 
our models.

Furthermore, It would be interesting to explore the circumstellar 
environments of those stars that are believed to have evolved 
from a RSG stage.
Some of those stars may still have observable relics
of the slow and dusty winds that they emitted 
when they were red supergiants.

\section*{Acknowledgments}
GM and HS thank Vincent Chomienne for having computed the first stellar 
models in the frame of this project.
CG acknowledges support from EU-FP7-ERC-2012-St Grant 306901.
We thank the anonymous referee for useful comments.

\appendix
\section{Rotation profiles in  BSG models}

\begin{figure}
\epsfig{file=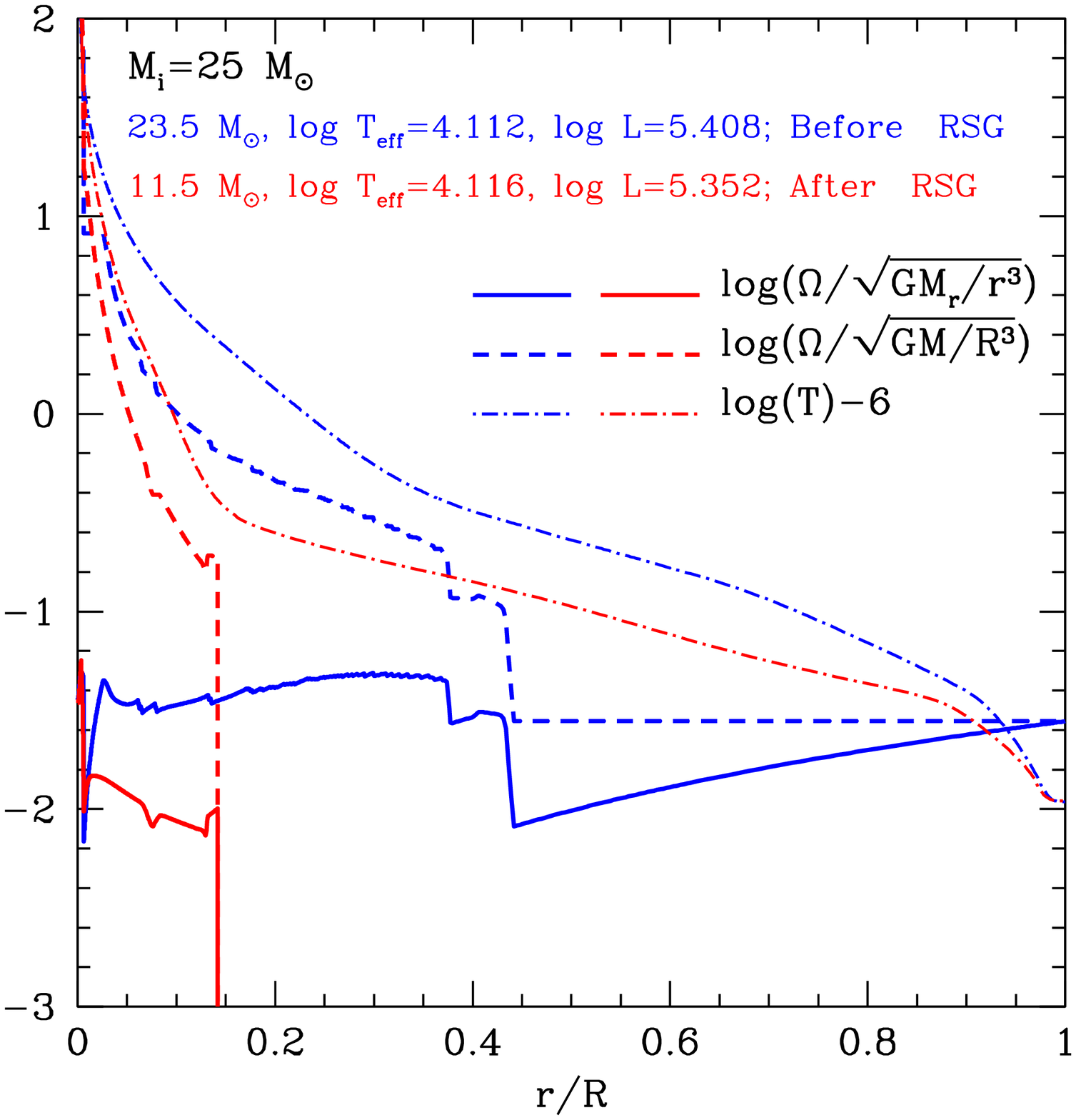,width=0.5\textwidth}
\caption{ Rotation and temperature profiles as functions
of the fractional radius, $r/R$, in two BSG models. 
The two models started with $M_{\rm i}=25M_\odot$, and have
similar effective temperatures, but are on different evolutionary stages.
Blue lines are used for the model evolving toward the red supergiant stage,
while red lines are for the model evolving on the blue-loop after the red 
supergiant stage.
The rotation frequency, $\Omega(r)$, is normalized in two different ways; 
by local gravity frequency $\sqrt{GM_r/r^3}$ (solid lines), and by
the global one $\sqrt{GM/R^3}$ (dashed lines). 
}
\label{fig:rot}
\end{figure}

For models including  rotation effects, the initial rotation speed is assumed 
to be 40\% of the critical rate at the surface of the zero-age main-sequence 
models.
Although the rotation speed is considerable during the main-sequence stage, 
it decreases significantly in the envelopes of supergiant models because of the
expansion and mass loss. 
Fig.\,\ref{fig:rot} shows runs of angular frequency of rotation in two 
BSG models having similar effective temperatures;
one (blue lines) is evolving toward the red supergiant stage and one 
(red lines) evolving on the blue loop after the red supergiant stage. 
In this figure, each rotation profile, $\Omega(r)$, is normalized by two 
different quantities; $\sqrt{GM_r/r^3}$\,(solid line) and 
$\sqrt{GM/R^3}$\,(dashed line).
The solid lines indicate that in both models the mechanical effect of rotation 
on the stellar structure is small because it is much smaller than the local 
gravity throughout the interior; i.e., $\Omega(r)/\sqrt{GM_r/r^3}\ll 1$.

The dashed lines in Fig.\,\ref{fig:rot} indicate rotation frequency 
itself normalized 
by the global parameters in the same way as the normalized pulsation frequency 
shown, e.g., in Fig.\,\ref{fig:omega}.
Although the rotation frequency is very high in the core of a supergiant, 
in the envelope it is much smaller than the pulsation frequencies; $\ga 1$ for 
radial pulsations (Fig.\,\ref{fig:omega}) and $\ga 0.3$ for nonradial 
oscillatory convection modes in the same normalization.
Since these pulsations are well confined to the stellar envelope 
($\log T < 6$, see Figs.\,\ref{fig:work3} and \ref{fig:nradwork}),
rotation hardly influence the pulsation property in these supergiant models.
This justifies our pulsation analyses without including 
the effect of rotation.

\bsp

\label{lastpage}

\end{document}